\documentclass[aps, prd, 10pt, twocolumn, eqsecnum, tightenlines, notitlepage, superscriptaddress, nofootinbib, preprintnumbers, floatfix]{revtex4-1}
\pdfoutput=1
\usepackage{epsfig,amsfonts,mathrsfs,amsmath,amssymb,graphicx,color,slashed}
\usepackage{amsmath,latexsym,amssymb,graphicx,slashed,hyperref,color,enumerate,url,etoolbox}
\hypersetup{colorlinks,citecolor= nicegreen,linkcolor= nicered}
\definecolor{nicered}{rgb}{0.7,0.1,0.1}
\definecolor{nicegreen}{rgb}{0.1,0.5,0.1}

\def\Caltech{Walter Burke Institute for Theoretical Physics, 
California Institute of Technology, Pasadena, CA 91125}
\def\Northwestern{Department of Physics and Astronomy,
Northwestern University, Evanston, IL 60208}
\def\Fermilab{Theoretical Physics Department, Fermilab, P.O. Box 500, Batavia, IL 60510}

\begin{document}

\title{Lepton Flavorful Fifth Force and Depth-dependent Neutrino Matter Interactions}

\author{Mark B.\ Wise}
\affiliation{\Caltech}

\author{Yue Zhang}
\affiliation{\Fermilab}
\affiliation{\Northwestern}

\date{\today}

\begin{abstract}

We consider a fifth force to be an interaction that couples to  matter with  a  strength that grows with the number of atoms. In addition to competing with the strength of gravity a fifth force can give rise to violations of the equivalence principle.  Current long range constraints on the strength and range of fifth forces are very impressive. Amongst possible fifth forces are those that couple to lepton flavorful charges  $L_e-L_{\mu}$ or $L_e-L_{\tau}$. They have the property  that their range and strength are also constrained by neutrino interactions with matter. In this brief note we review the existing constraints on  the allowed parameter space in gauged $U(1)_{L_e-L_{\mu}, L_{\tau}}$. We find two regions where neutrino oscillation experiments are at the frontier of  probing such a new force.  In particular, there is an allowed range of parameter space where  neutrino matter interactions relevant for long baseline oscillation experiments depend on the  depth of the neutrino beam below the surface of the earth.  
\end{abstract}

\preprint{CALT-TH-2018-012}
\preprint{FERMILAB-PUB-18-055-T}
\preprint{NUHEP-TH-18-04}

\maketitle

\section{Introduction}

%


One aspect of neutrino physics is that the interactions of neutrinos with matter play an important role in their propagation through  the sun and earth.  Constraints on   beyond the standard model sources  for  such interactions are usually  presented as limits on  the parameters $\epsilon _{ij}$ where $i$ and $j$ go over the neutrino flavors $e, {\mu}, \tau$. Long baseline neutrino interactions are sensitive to differences between the diagonal elements of the hermitian $ 3\times 3$ matrix $\epsilon_{ij}$ and its off diagonal elements. 

In this paper we consider models where a new massive $U(1)$ vector boson couples to a charge that is either the difference of lepton numbers $L_e-L_{\tau}$ or $L_e-L_{\mu}$. Then tree level exchange of the $Z'$ gauge boson gives contributions to the difference of diagonal elements of the $\epsilon$ matrix, $\epsilon_{ee}-\epsilon_{\tau \tau}$ or $\epsilon_{ee}-\epsilon_{\mu \mu}$. The symmetry generated by the  charges $L_e-L_{\tau}$ or $L_e-L_{\mu}$ is assumed to only be broken by neutrino masses. It is possible to  construct renormalizable  gauge theories that effectively realize this scenario at low energies but there is nothing compelling or even attractive about them and we will not dwell on this further. In this scenario the rates for charged lepton flavor changing processes like $\mu \rightarrow e \gamma$ and $\mu$ to $e$ conversion in the presence of a nucleus are suppressed by small neutrino masses and for the ranges of couplings and vector boson masses we consider have negligible rates \footnote{The new vector bosons will mediate at tree level transitions between different mass eigenstate neutrinos.}. Our goal in this short paper is to identify the regions of parameter space (coupling of the new force and mass of the new vector boson) where neutrino matter interactions provide the best constraints. We restrict our attention to $M_Z'>10^{-19} {\rm eV}$  (i.e., $1/M_{Z'}$  less than about 10 AU.) and then find two such regions. One where  $10^{-13} {\rm eV}> M_Z' >10^{-17}{\rm eV}$ and the other of much shorter range  and larger coupling  ($50\,{\rm MeV}<M_{Z'}< 300\,{\rm MeV}$).


\section{Constraints}

\begin{figure}[t]
\centerline{\includegraphics[width=0.48\textwidth]{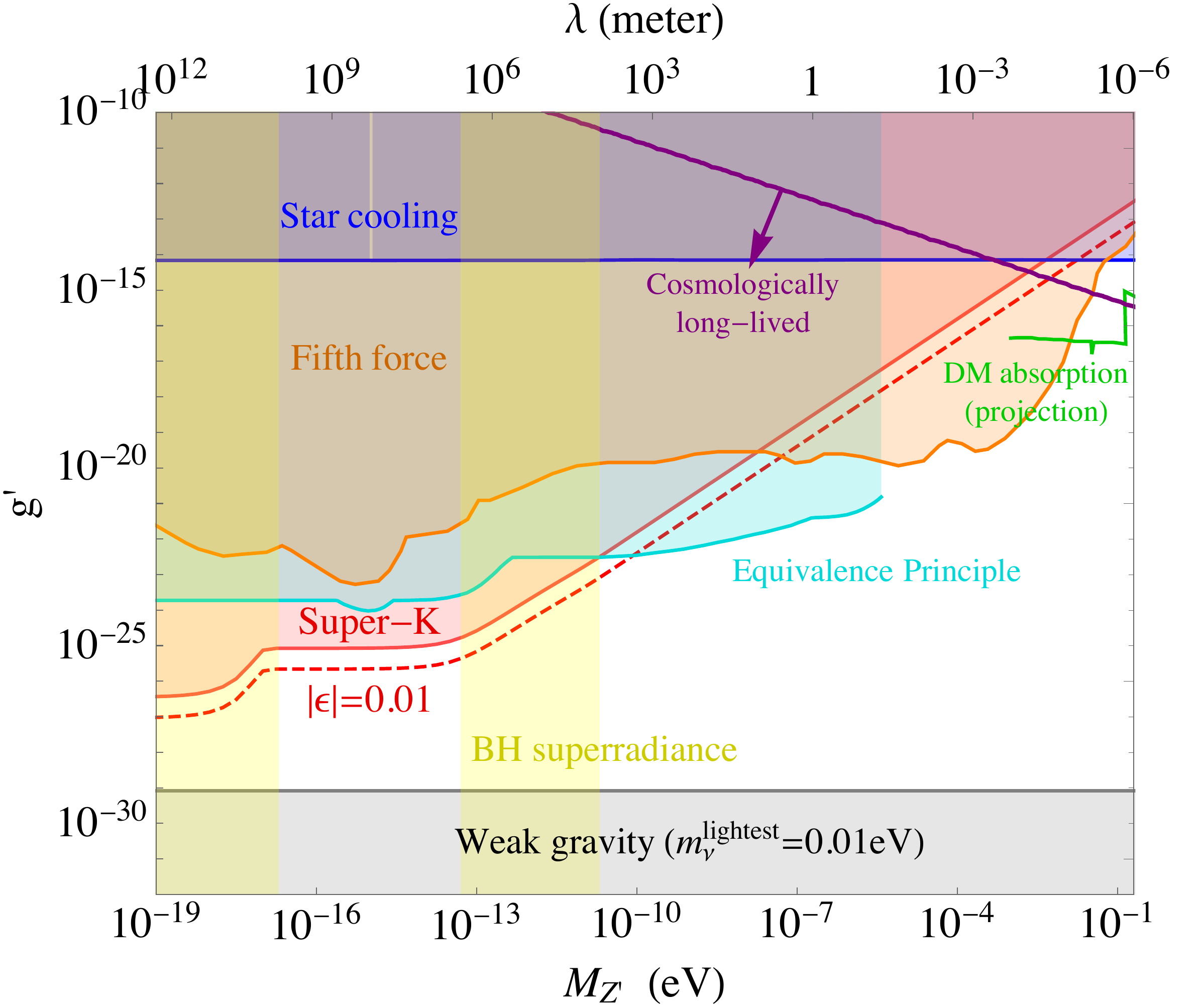}}
\caption{Constraints on the parameter space of a lepton flavorful long range force, which apply to gauge $L_e-L_\mu$ model (similarly for the $L_e-L_\tau$ model).
The solid (dashed) curve corresponds to a neutrino matter potential $\epsilon_{\mu\mu}=-\epsilon_{ee}=-0.147$ and $-0.01$, respectively.
The red shaded region has been excluded by Super-K.
The shaded orange (cyan) regions are excluded by fifth force (equivalence principle) test experiments.
The blue shaded region is excluded due to too much stellar cooling through the $Z'$ into neutrinos.
The gray shaded region is the lower limit on $g'$ by requiring there is at least one particle charged under the new $U(1)$ (assuming the lightest neutrino has mass 0.01\,eV) for which gravity is the weakest force. 
The yellow shaded region is ruled out by  superradiance using model dependent determinations of black hole spin.
Below the purple line, the $Z'$ is cosmologically long lived and could be the dark matter candidate, assuming the lightest neutrino is massless. If the lightest neutrino has a nonzero mass, the $g'$ coupling could be higher. 
The green curve shows the sensitivity of proposed direct detection experiments using special materials.
}\label{fig:longrange}
\end{figure}

We first focus on the window where $M_{Z'} <0.1\,{\rm eV}$, where the force can be  considered long range (compared with the distance between atoms in a typical solid). All the relevant constraints are summarized in Fig.~\ref{fig:longrange}.

\medskip
{\it Matter effect for neutrino oscillation.} \ \
In the presence of a long range $U(1)_{e-\mu}$ force, a muon neutrino traveling underground with depth $d$ feels an attractive potential energy from all the electrons around it  within a radius $\sim 1/M_{Z'}$. The potential created by all the electrons in the earth is
\begin{equation}\label{eq:Veff}
V_{\rm eff} = -2\pi \int_{0}^{\pi} \sin\theta  d\theta \int_0^{l_{\rm max}} l^2 \frac{n_e g'^2}{4\pi l} e^{-M_{Z'} l} d l \ ,
\end{equation}
where $l_{\rm max}(\theta) = (R_\oplus-d) \cos\theta + \sqrt{(R_\oplus-d)^2\cos^2\theta + (2R_\oplus-d)d}$, and $R_\oplus$ is the earth radius. 
Assuming the electron number density in the earth is a constant, the $l$ integral can be done analytically which yields
\begin{equation}
V_{\rm eff} = - \frac{n_e g'^2}{2M_{Z'}^2} \int_{0}^{\pi} \sin\theta  d\theta \left[ 1 - e^{- M_{Z'} l_{\rm max}} ( 1 + M_{Z'} l_{\rm max} ) \right] \ .
\end{equation}
We will do the $\theta$ integral numerically.
The potential energy felt by an electron neutrino is the opposite. In this case, there is no potential energy for the tau neutrino. Such flavor dependent matter potential could affect the splitting among (effective) neutrino masses and their oscillation probabilities~\cite{Wolfenstein:1977ue, Mikheev:1986gs}. The effective Hamiltonian due to new physics contribution to the matter potential is parametrized as
\begin{equation}
H_{\rm BSM} = \sum_{\alpha, \beta= e, \mu, \tau} \sqrt{2} G_F \epsilon_{\alpha\beta} (\bar \nu_\alpha \gamma^0 P_L \nu_\beta) n_e \ .
\end{equation}
In the $U(1)_{e-\mu}$ model, we have
\begin{equation}\label{eq:eps}
\epsilon_{\mu\mu} = - \epsilon_{ee} = \frac{V_{\rm eff}}{\sqrt2 G_F n_e} \ .
\end{equation}
In the massless $Z'$ limit, it has been realized that the matter effect in neutrino oscillations can be more sensitive to a small $g'$ than  fifth force and equivalence principle tests~\cite{Grifols:2003gy,Joshipura:2003jh}.  

In Fig.~\ref{fig:longrange}, the red curves are  constant\footnote{We take the underground neutrino to be at a depth $d=30\,$km, however, on the log-log plot the precise value of $d$ used is not relevant.} contours for $|\epsilon|$. There is a slope change near $M_{Z'}\simeq10^{-13}\,$eV where the potential energy sourced by the matter in the earth is most important and saturated, and
for $M_{Z'}$ below $10^{-17}\,{\rm eV} \sim (1\,{\rm AU})^{-1}$ there is another jump in the red curve where the potential energy sourced by electrons in the sun becomes more important.
The Super-K experiment set an upper limit on the matter effect based on atmospheric neutrino data $|\epsilon_{\mu\mu}-\epsilon_{\tau\tau}|<0.147$~\cite{Mitsuka:2011ty, Ohlsson:2012kf, Gonzalez-Garcia:2013usa} and has excluded the red shaded region. For reference, we also show the curve for $|\epsilon|\sim 0.01$ using the red dashed curve which perhaps may be probed by future neutrino experiments. The   shaded grey region which is labelled weak gravity is  where $g'^2 <G(m_{\nu}^{\rm lightest})^2$, when  $m_{\nu}^{\rm lightest}=0.01{\rm eV}$. 

\medskip
{\it Fifth force.} \ \
Fifth force experiments test the deviation from the $1/r$ gravitational potential between two objects. Denoting their atomic number by $Z_{1,2}$ and atomic weight by $A_{1,2}$, the potential energy for the leptonic $U(1)_{L_e-L_{\mu}, L_{\tau}}$ depends on the total number of electrons (related to $Z$) while gravitational energy depends on the total mass (related to $A$). Hence the total potential energy is
\begin{equation}
V(r) = - \frac{G m_1 m_2}{r} \left( 1 - \frac{g'^2 Z_1 Z_2}{4\pi G A_1 A_2 u^2} e^{-M_{Z'} r} \right) \ ,
\end{equation}
where $m_{1,2}$ are the mass of the two test objects, $G$ is the Newton's constant and $u\simeq 0.931\,$GeV is the atomic mass unit.
The quantity $\alpha_G\equiv ({g'^2 Z^2})/({G A^2 u^2})$ is bounded from above as a function of the interaction range $\lambda\equiv 1/M_{Z'}$~\cite{Kapner:2006si,Salumbides:2013dua}.
To translate the bound into that for $g'$ we make an approximation that for most materials $Z/A\simeq 0.5$.
The current limit on $g'$ is shown by the orange shaded region (excluded) in Fig.~\ref{fig:longrange}.

\medskip
{\it Equivalence principle.} \ \
The equivalence principle experiments test the non-universality of any long range force using two different materials (with $(Z_1, A_1)$ and $(Z_2,A_2)$) with equal total mass $m$ interacting with a common source with $(Z, A)$ and total number $M$. The difference in the leptonic $U(1)_{L_e-L_{\mu}, L_{\tau}}$ potential energy, normalized to the gravitational energy, takes the form
\begin{equation}
\frac{V_{1} (r) - V_{2} (r)}{V_G(r)}
= \frac{g'^2 Z}{4\pi G u^2 A}  \left(\frac{Z_1}{A_1} - \frac{Z_2}{A_2}\right) e^{-M_{Z'} r} \ .
\end{equation}
The quantity $\alpha\equiv ({g'^2 Z^2})({Z_1}/{A_1} - {Z_2}/{A_2})/({G A^2 u^2})$ bounded from above as a function of the interaction range $\lambda\equiv 1/M_{Z'}$~\cite{Schlamminger:2007ht}. The current limit on $g'$ is shown by the cyan shaded region (excluded) in Fig.~\ref{fig:longrange}.

In addition, there are also constraints on very light vector boson from the superradiance of spinning black holes~\cite{Baryakhtar:2017ngi}, which potentially exclude two mass windows shown by the yellow shaded bands in Fig.~\ref{fig:longrange}.  We will not strictly enforce these bounds when we discuss neutrino oscillations at DUNE and Super-K since they depend on models for black hole accretion disks \footnote{For a brief discussion of this issue see section {\it 10.1.2} of ~\cite{Teukolsky:2014vca}.}. 

In the same plot, below the purple curve, the $Z'$ can be cosmologically long lived against decaying into neutrinos and could be a dark matter candidate\footnote{Here we have assumed the lightest neutrino is massless. 
If the lightest neutrino is massive, the region for $Z'$ to be cosmological stable will be wider.
We do not address the issue of how the $Z'$ achieves it relic density. However, in that context it is worth noting that the coupling constants we consider are so small that the freeze in process only results in a very
small relic density.}.
There are proposed experiments to detect directly this type of dark matter by absorption using special materials~\cite{Hochberg:2017wce}.

For $Z'$ mass between $0.1\,{\rm eV}$ and the MeV scale, the stellar cooling gives the most important constraints~\cite{Jaeckel:2010ni}. 
For given stellar object with temperature well above $M_{Z'}$, the cooling bound on $g'$ is insensitive to $M_{Z'}$ because the cooling could occur via off-shell $Z'$ to neutrinos.

\begin{figure}[t]
\centerline{\includegraphics[width=0.48\textwidth]{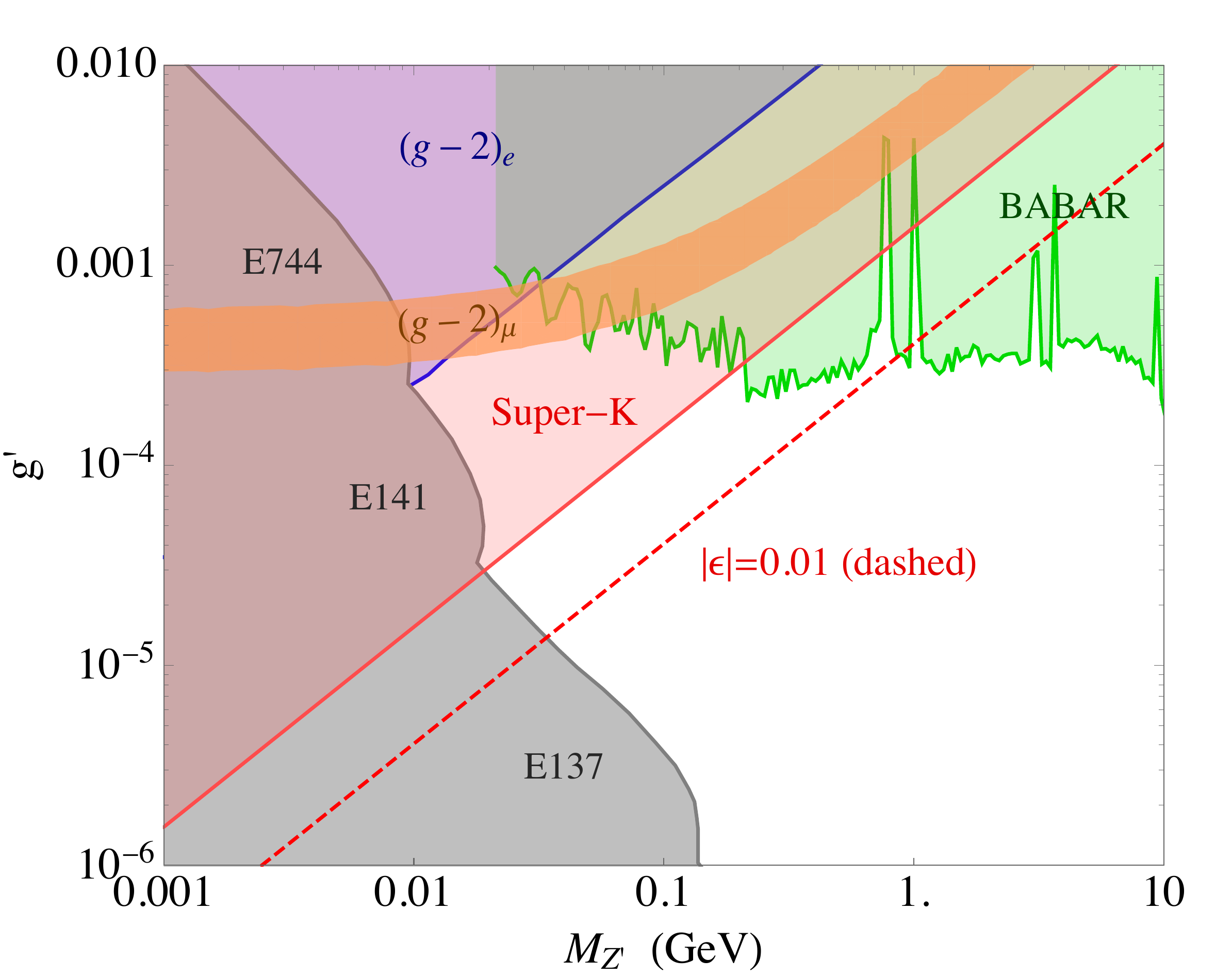}}
\caption{Constraints on the parameter space of a lepton flavorful short range force, which apply to gauge $L_e-L_\mu$ model (similarly for the $L_e-L_\tau$ model).
The BaBar, electron and muon $g-2$ and beam dump (E744, E141, E137) experimental constraints are translated from those for the dark photon.
The solid (dashed) curve corresponds to a neutrino matter potential $\epsilon_{\mu\mu}=-\epsilon_{ee}=-0.147$ and $-0.01$, respectively.
The red shaded region has been excluded by Super-K. This is another window where neutrino experiments stand at the frontier of probing such a new force.
}\label{fig:shortrange}
\end{figure}

Another window we examine in detail is for $M_{Z'}$ above an MeV, as shown in Fig.~\ref{fig:shortrange}.
Most of the constraints in this plot are translated from those for the dark photon searches~\cite{Essig:2013lka}.
The green, blue and gray regions are excluded by BaBar , electron $g-2$ and electron beam dump experiments.
Unlike the dark photon, our $Z'$ also has a branching ratio to neutrinos which affect the branching ratio of its visible decay. We take this into account when translating the limits.
Interestingly, we find that there is a mass window, with $10\,{\rm MeV}< M_{Z'} <200\,{\rm MeV}$, where the neutrino experiments can do better than other experiments.
This point was made in a similar model in~\cite{Babu:2017olk}.

For $Z'$ heavier than the weak scale, the LEP constraint on contact interactions is found to be the strongest~\cite{Wise:2014oea}. 

\section{Depth-dependent matter effects}

\begin{figure*}[t]
\centerline{\hspace{3.4cm}\includegraphics[width=0.75\textwidth]{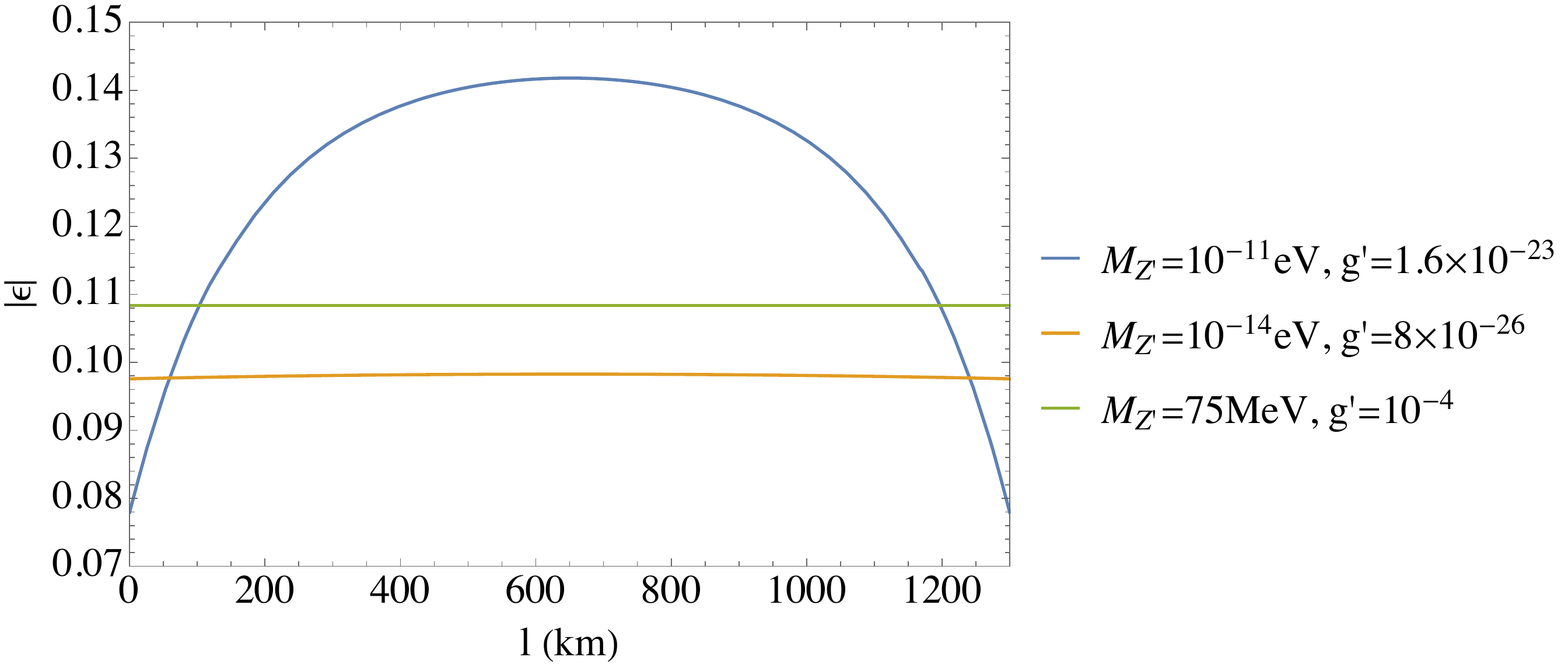}}
\caption{Neutrino matter effect in gauged $L_e-L_\ell$ ($\ell=\mu, \tau$) model for three choices of $g'$ and $M_{Z'}$ values.
We consider a similar setup to DUNE, where neutrino travels 1300\,km underground. 
For simplicity we assume both the source and the detector of neutrinos are located at the surface of the earth.
In this case the largest depth of neutrino reaches is $d_{\rm max}\simeq 33\,$km,
$|\epsilon|$ is plotted as a function of the distance $l$ that neutrino has traveled.
For the very short (long) range force case with $M_{Z'}\gg d_{\rm max}^{-1}$ ($M_{Z'}\ll d_{\rm max}^{-1}$), 
we find the $|\epsilon|$ is rough a constant as shown by the green (orange) curves; 
While for the intermediate range with $M_{Z'}\sim d_{\rm max}^{-1}$, 
the variation of the matter potential can be significant, as shown by the blue curve.
}\label{fig:compare}
\end{figure*}

The matter effect in neutrino oscillations could occur with either short- or long-range $Z'$ exchange. For a neutrino traveling through rocks, the contribution to matter potential from the new lepton flavored $U(1)$ interaction is given by Eqs.~(\ref{eq:Veff}) to (\ref{eq:eps}). First, we discuss the asymptotic behaviors of the $\epsilon$'s (assuming gauged $U(1)_{L_e-L_{\mu}}$)
\begin{equation}
\begin{split}
&\epsilon_{\mu\mu} = - \epsilon_{ee} = \frac{V_{\rm eff}}{\sqrt2 G_F} \\
&=\left\{
\begin{array}{ll}
- \frac{g'^2}{\sqrt2 G_F M_{Z'}^2}, & \hspace{0.5cm} M_{Z'}^{-1}\ll d, R_\oplus \\
- \frac{g'^2 R_\oplus^2}{3\sqrt2 G_F} \left( 1 + \frac{d}{R_\oplus} - \frac{d^2}{2R_\oplus^2} \right), & \hspace{0.5cm} M_{Z'}^{-1}\gg d, R_\oplus
\end{array}
\right.
\end{split}
\end{equation}
where $R_\oplus$ is the earth radius and $d$ is the depth of neutrinos from the earth surface. Here we assume the electron number density is uniform throughout the earth.
Assuming the earth is an iron ball, then the electron number density is $n_e \simeq 2.2\times10^{24}\,{\rm cm^{-3}}$.
In both the situations $M_{Z'}^{-1}\ll d, R_\oplus$ and  $M_{Z'}^{-1}\gg d, R_\oplus$, as long as $R_\oplus\gg d$, the matter effect is very insensitive to $d$.


The depth dependence can become important for intermediate range of the force where  the Compton wavelength of the $Z'$ is comparable to the depth.
In this case, there is a significant difference in the matter potential for a neutrino underground a depth $d=M_{Z'}^{-1}$ and one at the surface of the earth.
In the former case, all the space within a sphere of radius $M_{Z'}^{-1}$ is filled with electrons that source the new matter potential; while in the latter case,
only half of the sphere is filled. Therefore, the matter effect could vary by as much as a factor of 2.

For a long baseline experiment where the neutrino beam goes in a straight line the one point near the surface  to another where it is detected the relation between the depth to the surface $d$ and the distance travelled underground $l$ is, 
\begin{equation}
\label{dl}
d(l)=R_\oplus-\sqrt{R_\oplus^2+l^2-l l_{\rm tot}} \simeq {l(l_{\rm tot}-l) \over 2 R_\oplus},
\end{equation}
where $l_{\rm tot}$ is the total underground distance that the neutrino beam will travel between injection and detection. The expression on the far right of eq.~(\ref{dl}) is appropriate when $R \gg l_{\rm tot}$. The maximum depth of the beam below the surface of the earth, $d_{\rm max}$, occurs when the neutrino has traveled a distance $l=l_{\rm tot}/2$ and so,
$d_{\rm max} = R_\oplus-\sqrt{R_\oplus^2-l_{\rm tot}^2/4} \simeq l_{\rm tot}^2/(8R_\oplus)$. In Fig.~\ref{fig:compare}, we plot the $l$ dependence of $|\epsilon|$, for $l_{\rm tot}=1300~{\rm km}$ using three different values of $g'$ and $M_{Z'}$.

\begin{figure*}[t]
\centerline{\includegraphics[width=0.49\textwidth]{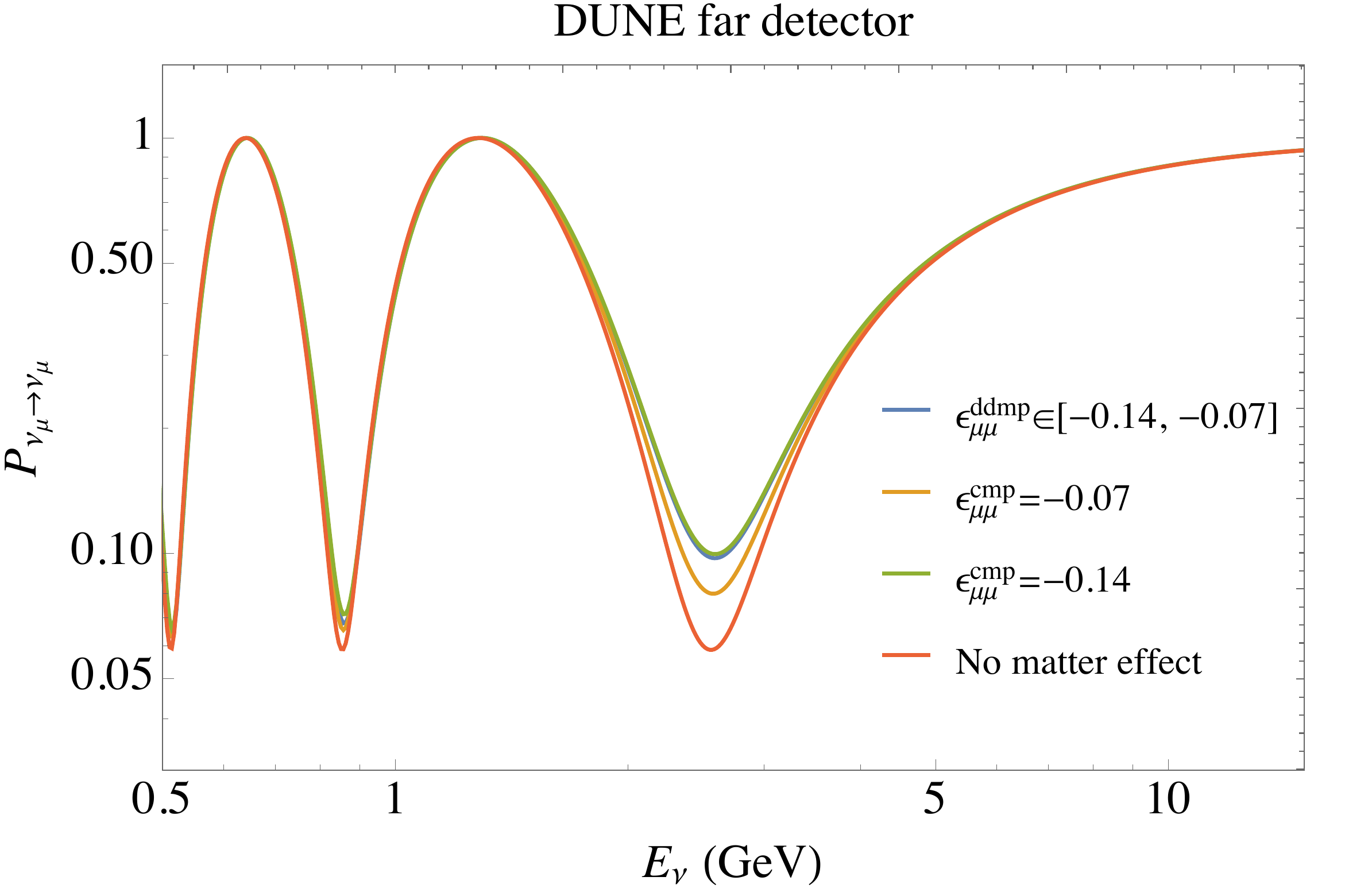}\includegraphics[width=0.5\textwidth]{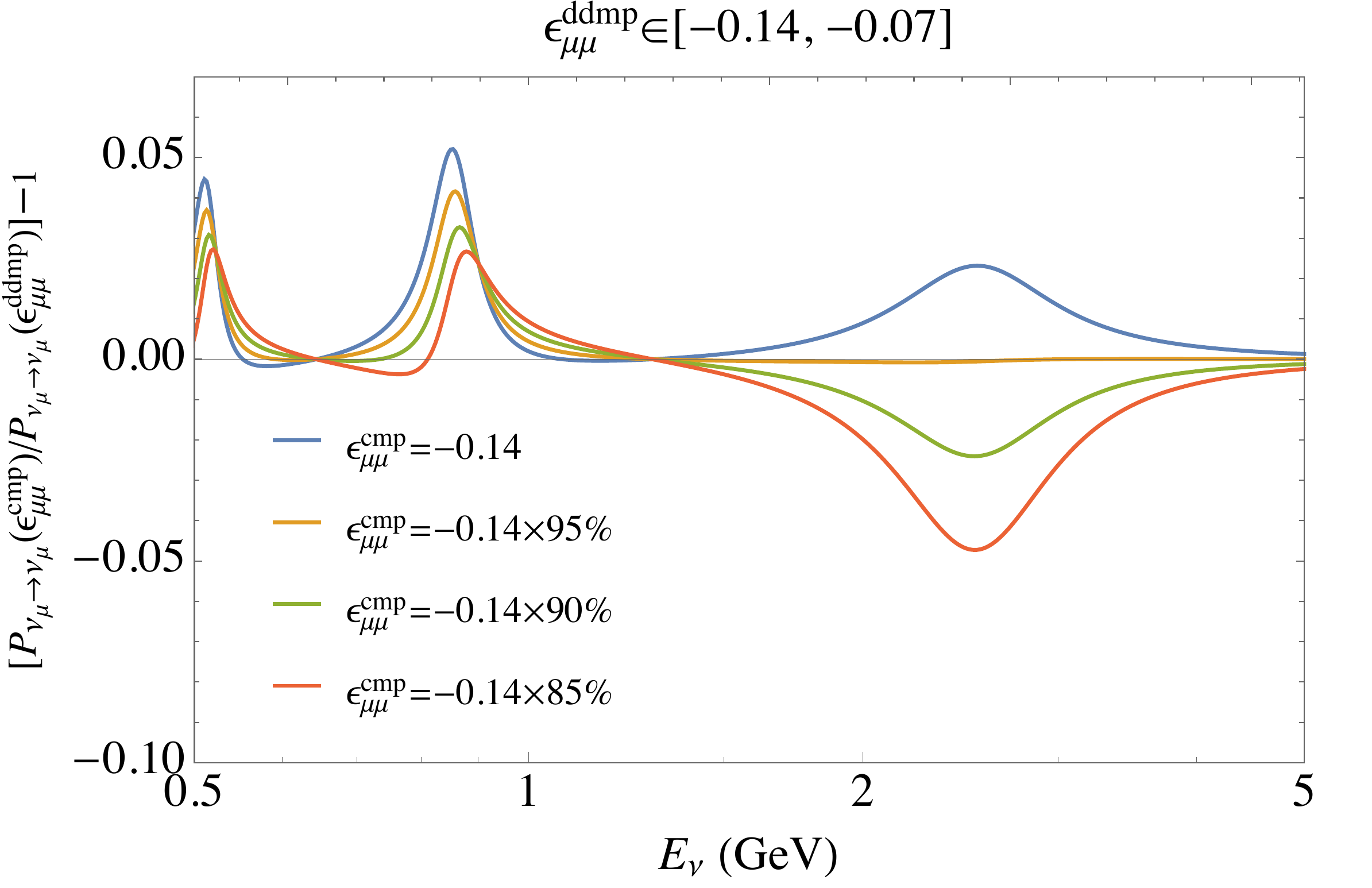}}
\caption{{\sf Left:} Survival probability of a muon neutrino beam after traveling through the earth in a DUNE-like setup as discussed in the caption of Fig.~\ref{fig:compare}.
We consider a gauged $L_e-L_\mu$ model and simplified two flavor oscillations $\nu_\mu \to \nu_\tau$.
{\sf Right:} A closer comparison between a depth dependent matter potentials defined in Eq.~(\ref{ddmp}) and several constant matter potentials.
\vspace{0.2cm}}\label{fig:DUNE}
\centerline{\includegraphics[width=0.48\textwidth]{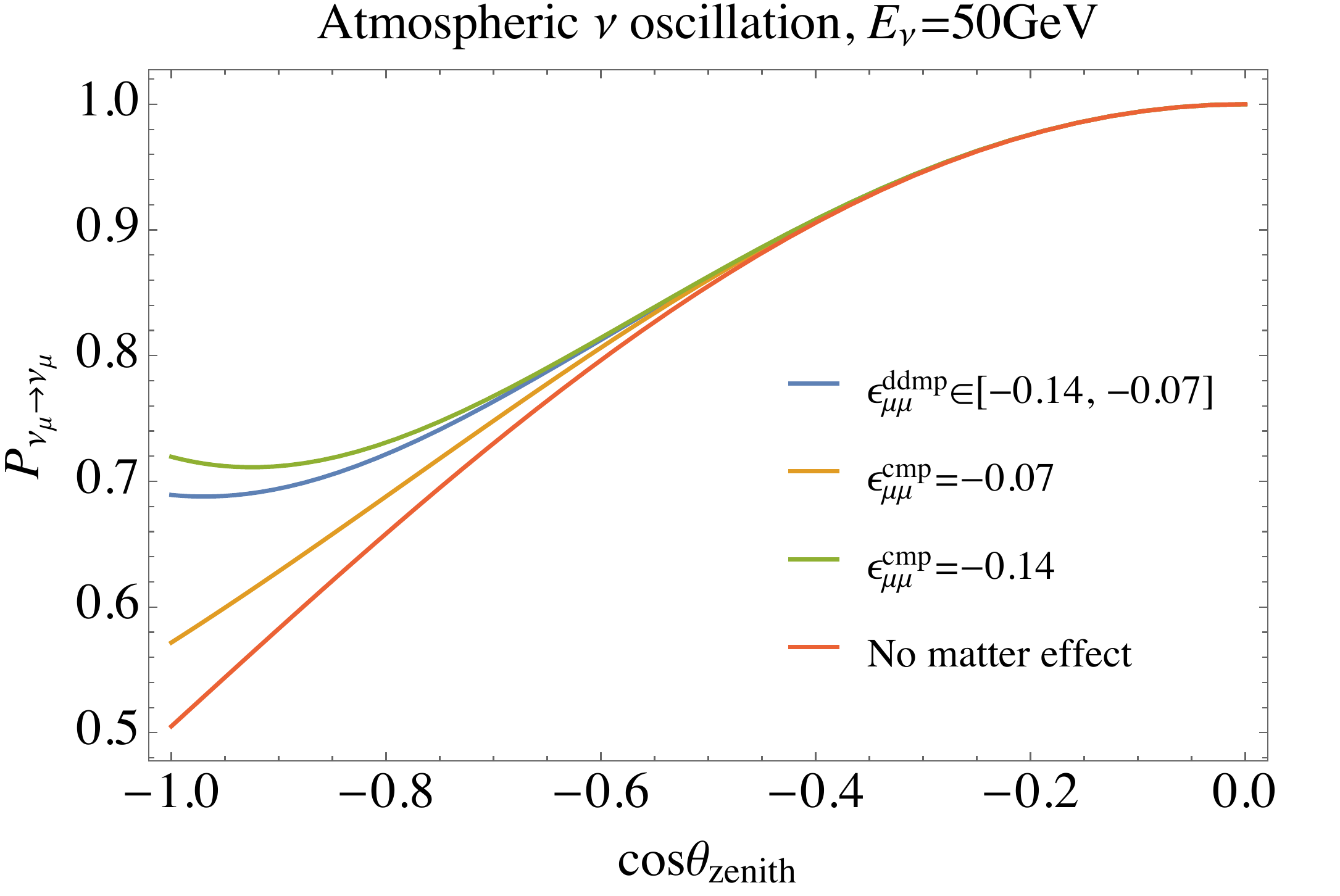}\includegraphics[width=0.5\textwidth]{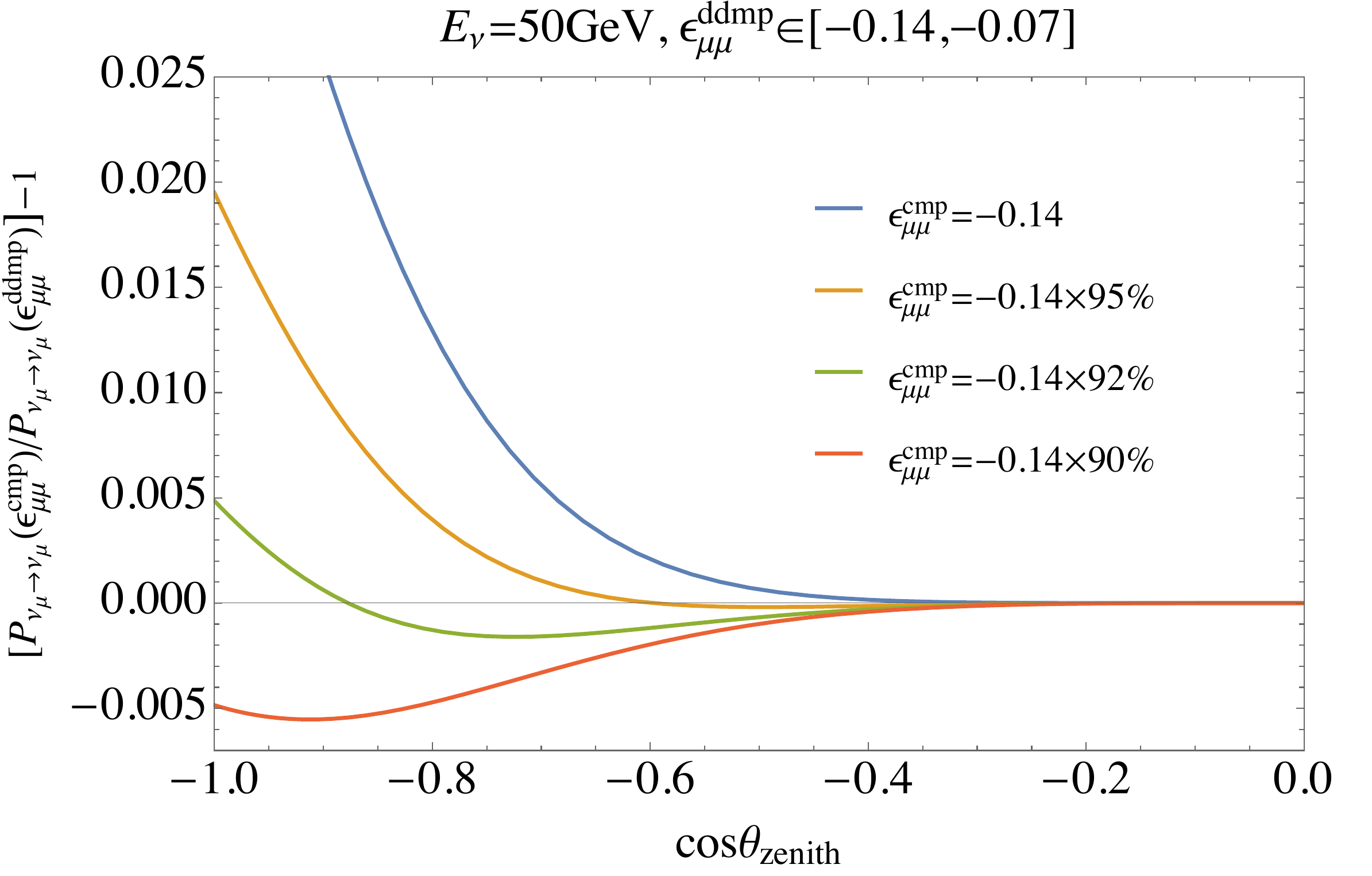}}
\caption{{\sf Left:} Survival probability of an atmospheric muon neutrino after traveling through the earth and the zenith angle dependence in a gauged $L_e-L_\mu$ model.
We consider simplified two flavor oscillations $\nu_\mu \to \nu_\tau$. 
The blue curve corresponds to a depth dependent matter potential, while the orange and green curves corresponds to constant matter potentials.
The red curve is the SM limit muon and tau neutrinos experience identical matter potential.
{\sf Right:} A closer comparison between a depth dependent and several constant matter potentials.}\label{fig:ATMO}
\end{figure*}

In Fig.~\ref{fig:DUNE} (left), we plot the survival probability of a muon neutrino beam as a function of the neutrino energy, after traveling $l_{\rm tot}=1300\,$km distance, in the presence of earth matter effect due to a $Z'$ mediated $L_e-L_\mu$ fifth force. 
The features for antineutrino oscillations are similar, expect that the matter potential has an opposite sign.
For simplicity, we consider two flavor oscillations between $\nu_\mu$ and $\nu_\tau$.
The vacuum mass square difference and mixing angles are chosen to be $\Delta m_{23}^2=2.44\times10^{-3}$ and $\theta_{23}=38^\circ$. 
Here we find that choosing the value of $\theta_{23}$ to be away from maximal mixing allows the matter effects to be more visible.
The blue curve corresponds to a depth dependent long range force with $M_{Z'}= 10^{-11}\,$eV\,$\simeq (20\,{\rm km})^{-1}$. We choose the value of $g'=1.6\times10^{-23}$ such that the largest NSI effect (which occurs when the neutrino is half way through, with $d_{\rm max}\simeq 33\,$km) is equal to $\epsilon_{\mu\mu} \simeq 0.14$.
In this case when the neutrino is on the surface, $\epsilon_{\mu\mu}\simeq 0.07$. We denote such a depth dependent matter potential (ddmp) with 
\begin{equation}\label{ddmp}
-0.14 < \epsilon_{\mu\mu}^{\rm ddmp} < -0.07 \ ,
\end{equation}
which is a function of $l$.
In contrast, the orange and green curves correspond to two limiting cases with constant matter potential (cmp), 
\begin{equation}
\epsilon_{\mu\mu}^{\rm cmp}\simeq -0.07\ \ {\rm and}\ \ -0.14 \ , 
\end{equation}
respectively.
Clearly, the result for depth dependent case lies in between the two limiting cases and all the curves have similar shapes. 
Naively, one might expect that the oscillation probability due to a depth dependent matter potential $\epsilon_{\mu\mu}^{\rm ddmp}$ could be mimicked with a proper choice of constant matter potential $\epsilon_{\mu\mu}^{\rm cmp}$.
In Fig.~\ref{fig:DUNE} (right), we plot the ratio of $P_{\nu_\mu\to\nu_\mu}(\epsilon_{\mu\mu}^{\rm cmp})$ to $P_{\nu_\mu\to\nu_\mu}(\epsilon_{\mu\mu}^{\rm ddmp})$ minus 1, for several choices of $\epsilon_{\mu\mu}^{\rm cmp}$ but with $\epsilon_{\mu\mu}^{\rm ddmp}$ held the same as defined above.
Interestingly, although for fixed neutrino energy one can adjust $\epsilon_{\mu\mu}^{\rm cmp}$ so that the two probabilities are made the same, 
this is not possible for all the energies. Generically, the difference is at a few percent level.
For $E_\nu\simeq0.5$ and 0.85\,GeV, the difference can be a difference as large as 5\%.
It seems very challenging for DUNE to distinguish a depth dependent matter potential based on the $P_{\nu_\mu\to\nu_\mu} (E_\nu)$ spectrum, 
but it may be possible to do so at future experiments with higher precision.

As another example, we consider the matter effect on zenith angle dependence in atmospheric neutrino oscillations. In this case, we choose a smaller $Z'$ mass, $M_{Z'}=3\times 10^{-14}\,{\rm eV} \simeq R_\oplus^{-1}$. 
Here we choose the value $g'=9.1\times10^{-26}$ so that for a neutrino at the center of the earth we have $\epsilon_{\mu\mu} \simeq - 0.14$.
In Fig.~\ref{fig:ATMO} (left), we plot the muon (anti)neutrino survival probability for zenith angle $\theta$ between $\pi/2$ to $\pi$. In this range, neutrinos must travel through the earth to reach the detector. The depth of neutrino underground after traveling a distance $l$ is still given by Eq.~(\ref{dl}) with $l_{\rm tot}= 2 R_\oplus \cos(\pi-\theta_{\rm zenith})$.
For simplicity, we neglect the part of neutrino path that is through the atmosphere.
The atmospheric neutrinos can have much higher energy than the accelerator neutrinos, thus we choose $E_\nu=50\,$GeV.
As a result, the matter effect is much more visible.
We choose the vacuum oscillation parameters to be $\Delta m_{23}^2=2.44\times10^{-3}$, and $\theta_{23}=45^\circ$ this time.
The depth dependence matter effect still does not give a very different shape than the constant matter effect cases, but the survival probability differences at $\theta=\pi$ are a lot more visible in this case. 
In Fig.~\ref{fig:ATMO} (right), we again compare the oscillation probabilities due to the depth-dependent matter potential as mentioned above versus a constant one.
Similar to the DUNE case, the former cannot be completely mimicked at every zenith angle by the latter, but the difference in this case can be as small as half a percent.

With extremely high precision neutrino oscillation data one might be able to find some evidence for a depth dependent matter effect.
One would also need to distinguish the depth dependence due to the range of the Yukawa potential from variations of earth matter density along the path.

\section{Concluding Remarks}

We considered neutrino matter interactions that arise from gauging $U(1)_{L_e-L_{\mu}}$ or $U(1)_{L_e- L_{\tau}}$. Restricting our attention to $M_Z' > 10^{-19}{\rm eV}$ we found two regions of $Z'$ mass and coupling constant where  the impact of neutrino interactions  with matter provide the best constraints. In these regions future long baseline neutrino experiments may improve the bounds. For a range of $Z'$ masses neutrino matter interactions depend on the depth of neutrinos below the surface of the earth. 
We discussed in some detail the impact on neutrino oscillation probabilities as well as how it may be distinguished from the case with a constant matter potential. 

In this paper we neglect the kinetic mixing between the gauge groups $U(1)_Y$ and $U(1)_{L_e-L_{\mu}, L_{\tau}}$. The dimensionless parameter that characterizes this mixing is naturally of order $g' e/(16\pi)^2$. For the small values of $g'$ considered in this paper including it at this level  would not introduce further constraints.

Finally, we note that similar results hold for gauged,  $U(1)_{aB+bL_e+cL_{\mu}+c L_{\tau}}$, which is anomaly free if $3a+b+c+d=0$.

\section*{Acknowledgement}
We would like to thank Andr\'e de Gouv\^ea, Stephen Parke and Saul Teukolsky for helpful discussions.
The work of MBW was supported by the DOE Grant DE-SC0011632 and by the Walter Burke Institute for Theoretical Physics.
The work of YZ was supported in part by DOE Grant DE-SC0010143.  
This manuscript has been authored by Fermi Research Alliance, LLC under Contract No.~DE-AC02-07CH11359 with the U.S. Department of Energy, Office of Science, Office of High Energy Physics. The United States Government retains and the publisher, by accepting the article for publication, acknowledges that the United States Government retains a non-exclusive, paid-up, irrevocable, world-wide license to publish or reproduce the published form of this manuscript, or allow others to do so, for United States Government purposes.

\end{document}